\documentclass[12pt]{article}
\usepackage{graphicx}

\newcommand\pubnumber{WSU--HEP--XXYY}
\newcommand\pubdate{\today}

\def\support{\footnote{Work supported by the Slovenian Research Agency.}}

\def\Title#1{\begin{center} {\Large #1 } \end{center}}
\def\Author#1{\begin{center}{ \sc #1} \end{center}}
\def\Address#1{\begin{center}{ \it #1} \end{center}}

\newcommand\pubblock{\rightline{\begin{tabular}{l} \pubnumber\\
         \pubdate  \end{tabular}}}
\newenvironment{Abstract}{\begin{quotation}  }{\end{quotation}}
\newenvironment{Presented}{\begin{quotation} \begin{center} 
             PRESENTED AT\end{center}\bigskip 
      \begin{center}\begin{large}}{\end{large}\end{center} \end{quotation}}

\def\beq{\begin{equation}}
\def\eeq#1{\label{#1}\end{equation}}
\def\eeqn{\end{equation}}

\def\beqa{\begin{eqnarray}}
\def\eeqa#1{\label{#1}\end{eqnarray}}
\def\eeqan{\end{eqnarray}}

\let\bar=\overbar

\def\Dslash{\not{\hbox{\kern-4pt $D$}}}
\def\dslash{\not{\hbox{\kern-2pt $\del$}}}

\def\msb{{\bar{\ssstyle M \kern -1pt S}}}

\newcommand{\qq}{q^2}

\newcommand {\E}[1]{\times 10^{#1}} 
\newcommand{\mc}[1]{\mathcal{#1}}

\newcommand{\mrm}[1]{\mathrm{#1}}

\renewcommand{\Im}[0]{\mrm{Im}}

\begin{document}
\begin{titlepage}
\pubblock

\vfill
\Title{Theoretical perspective on rare and radiative charm decays }
\vfill
\Author{Svjetlana Fajfer\support}
\Address{Department of Physics, University of Ljubljana, Jadranska 19, 1000 Ljubljana, Slovenia\\
and J. Stefan Institute, Jamova 39, P. O. Box 3000, 1001 Ljubljana, Slovenia \\
}
\vfill
\begin{Abstract}
Recent  experimental bounds on rare charm decays offer  a chance to improve our theoretical understanding of physics present in $c \to u \gamma$ and $c \to u l^+ l^-$ transitions.
Standard Model and New Physics contributions are reviewed for inclusive and exclusive $D \to V\gamma$, $D^+ \to \pi^+ l^+ l^-$,  $D \to l^+ l^-$ decays. Observables important for search of New Physics are discussed. Possibility to observe  CP violation in rare charm decays is questioned.
\end{Abstract}
\vfill
\begin{Presented}
The 7th International Workshop on Charm Physics (CHARM 2015)\\
Detroit, MI, 18-22 May, 2015
\end{Presented}
\vfill
\end{titlepage}
\def\thefootnote{\fnsymbol{footnote}}
\setcounter{footnote}{0}
%


\section{Introduction}
At low energies  New Physics (NP) was expected to be seen indirectly in the down quark sector. The LHC offered chance for direct search of NP.
Although, NP is not found yet directly at high energies, there are a number of reasons why we still expect to find its presence.
For example on experimental side,  in B physics  tensions between SM expectations and experimental results are found. 
It was noticed that in $B \to K^* \mu^+ \mu^-$ observable, known as $P_5^\prime$,  deviates for about $3 \sigma$  \cite{Descotes-Genon:2013vna} from the SM prediction, the ratio $R_{D*}^{\tau,l}$  
$={ BR (B \to D \tau \nu_\tau)}/{ BR (B \to KD \mu \nu_\mu)}$ exhibit $3.8 \sigma$ deviations \cite{Fajfer:2012vx} and $R_K= { BR (B \to K \mu^+ \mu^-)_{q^2 \in [1,6] \rm{GeV}^2}}/{ BR (B \to K e^+ e^-)_{q^2 \in [1,6] \rm{GeV}^2}}$ has $2.6 \sigma$ discrepancy from the SM value \cite{Hiller:2014yaa}. 
Two of these observables  in $B\to K^{(*)}$ transitions are result of flavor changing neutral current processes (FCNC), while the one for $B\to D^{(*)}$  is a result of charge current. 
These anomalous results stimulated numerous studies of NP in B meson system.
One should keep in mind that all other B meson physics observables offer additional constraints to new physics.\\
Top quark physics seems to be important for NP searches in the up-like quark sector. Properties  and dynamics of top quarks attract a lot of attention on experimental and theoretical side.  However, there is no indication yet about NP presence.  
The  question one can ask: is there any chance to observe NP effects in  charm FCNC physics? The constraints on NP in semileptonic charm decays driven by charge currents have been discussed in Ref. \cite{Fajfer:2015ixa}. 
On the other hand, FCNC rare charm processes are  accessible in radiative or semileptonic decays in which transitions $c \to u \gamma$ and $c \to u l^+ l^-$ occur.
The main obstacle to search for NP in rare charm decay is the presence of many non-charm resonances in the vicinity of D mesons masses. 
Strong role of GIM mechanism is very important in charm FCNC dynamics. The interplay of CKM parameters and masses of down-like quarks leads to strong suppression of all  FCNC in  D  meson processes. In addition, long distance contributions overshadow short distance effects. The main issue is how to separate  information on short distance dynamics, either within SM or in its extensions. This is  a longstanding problem in rare charm decays.\\
Three years ago  flavor community was concerned about 
discrepancy between measured and expected CP violating asymmetry in charm decays \cite{Aaij:2011in}. 
Although this discrepancy seems to disappear,  many studies  and additional checks of the observed anomaly in rare charm decays were performed.  The question 
 on observability  of CP violation in charm rare decays, should be answered.
In Sec. 2  contributions to  $c \to u \gamma$ and $c \to u l^+ l^-$  decay modes are  reviewed. The exclusive weak radiative $D\to V \gamma$  decays  are discussed in Sec. 3. and $D\to \mu^+ \mu^-$, $D\to P(P') \mu^+ \mu^-$ were analysed in Sec. 4.  Tests of CP violation in charm meson decays with the leptons in the final state  are discussed in Sec. 5. Last section  contains  the summary.

\section{Inclusive decay modes: $c \to u \gamma$ and $c \to u l^+ l^-$ }

The $c \to u \gamma$ and $c \to u l^+ l^-$  transitions within SM  can be approached by the effective low-energy Lagrangian:
\begin{equation}
{\cal L}^{SD}_{eff} = -\frac{4 G_F}{\sqrt 2} V_{cb}^* V_{ub} \sum_{i=7,9,10} C_i Q_i,
\label{e1}
\end{equation}
The operators are then:
\begin{eqnarray}
Q_7 &=&\frac{e}{8 \pi^2} m_c F_{\mu \nu} \bar u \sigma_{\mu \nu}(1+\gamma_5) c,\nonumber\\
Q_9 &=&\frac{e^2}{16 \pi^2} \bar u_L \gamma_{\mu } c_L \bar l \gamma^\mu l,\nonumber\\
Q_{10} &=&\frac{e^2}{16 \pi^2} \bar u_L \gamma_{\mu } c_L \bar l \gamma^\mu \gamma_5 l.\label{e2}
\end{eqnarray}
In (\ref{e1}) $C_i$ denote as usual Wilson coefficients  (they are determined at the scale $\mu = m_c$), $F_{\mu \nu}$ is the electromagnetic field strenght and $q_L = \frac{1}{2} ( 1-\gamma_5) q$. 
In the case of $c \to u \gamma$ decay only $C_7$ contributes, while in the case of $c \to u l^+ l^-$ all three Wilson coefficents are present. 
 The QCD corrections enhance the rate to  $BR(c \to u \gamma )_{SM} =2.5 \times  10^{-8}$ ~\cite{Greub:1996wn,HoKim:1999bs}. Within Standard model the short distance contribution coming from $Q_{7,9}$  leads to the branching ratio 
$BR(D\to X_u e^+ e^-)^{SD}_{SM} \simeq 3.7 \times 10^{-9}$  \cite{Burdman:2001tf,Fajfer:2001sa,Paul:2011ar}.  
Long distance contributions  overshadow the short distance one with 
$BR(D\to X_u e^+ e^-)^{LD}_{SM} \sim {\cal O}(10^{-6})$ \cite{Burdman:2001tf,Fajfer:2001sa}.

\section{Exclusive decay modes: $D \to V \gamma$ } 

Previous studies of the these decays were based on the knowledge of non-leptonic weak decays of charm mesons to two light vectors and then vector meson dominance was assumed to predict rates for $D \to V \gamma$ \cite{Burdman:1995te},  or  a model of charm mesons as heavy mesons accompanied by hidden symmetry approach for the vector mesons as done in \cite{Fajfer:1998dv}. There are also QCD sum rules calculation done by  authors of Ref. \cite{Khodjamirian:1995uc} and more recent one in Ref. \cite{Lyon:2012fk}. It was found that the amplitudes fulfil relations ${\cal A} (D^0 \to \bar K^{*0}\gamma) \simeq {\cal A} (D^0 \to  \rho^{0}\gamma)$ and ${\cal A} (D^+_s \to \rho^{+}\gamma) \simeq {\cal A} (D^+ \to \rho^{+}\gamma)$  \cite{Khodjamirian:1995uc} that led to the predictions for the branching ratios 
$BR(D^0 \to \bar K^{*0}\gamma) \simeq 1.5\times 10^{-4}$, $BR(D^0 \to \rho^{0}\gamma) \simeq 3.1\times 10^{-6}$, $BR (D^+_s \to \rho^{+}\gamma) \simeq 2.8\times10^{-5}$ and $BR(D^+ \to \rho^{+}\gamma) \simeq 2.7 \times 10^{-6}$. On the experimental side there only two results for the branching ratios 
$BR(D^0 \to \bar K^{*0}\gamma)_{exp} \simeq 3.27(34) \times10^{-4}$ and $BR(D^0 \to \phi \gamma)_{exp} \simeq 2.70(35)  \times10^{-5} $. 
One might update  calculation of  \cite{Burdman:1995te} including more recent results  for the $D\to V_1 V_2$  helicity amplitudes. However, the relative phases of different contributions are still not possible to obtain and only range of the values for the branching ratios can be given: 
$BR(D^0 \to \bar K^{*0}\gamma) \simeq (2.8-4.9) \times10^{-4} $ and $BR(D^0 \to \phi \gamma) \simeq  (2.8-4.1)\times10^{-5} $.

\section{   $D^0 \to \mu^+ \mu^-$, $D\to P \mu^+ \mu^-$ and  $D\to P_1 P_2 \,\mu^+ \mu^-$}

The LHCb collaboration improved  the  bound on  the rate $BR(D^0 \to \mu^+ \mu^-) < 6.2 (7.6) \times 10^{-9}$  \cite{Aaij:2013cza} and  for the first time, they 
determined limits  on the branching fractions in several di lepton invariant mass bins  in $BR(D^+ \to \pi^+ \mu^+ \mu^-) < 7.3(8.3) \times 10^{-8}$  \cite{Aaij:2013sua}.  At the low  dilepton invariant  mass region  $0.25$ GeV $\le m_{\mu \mu}  \le 0.525$ GeV the LHCb collaboration found upper bound on the rate  $BR( D^0 \to \mu^+ \mu^-)_{l.e.b} < 2 \times 10^{-8}$, while at high dilepton invariant mass  $1.25$ GeV $\le m_{\mu \mu}  \le 2.0$ GeV,   $BR_{h.e.b}( D^0 \to \mu^+ \mu^-) < 2.6 \times 10^{-8}$   \cite{Aaij:2013sua}, at $90\%$ confidence level.  
These two results enable to constrain size of the Wilson coefficients entering effective Lagrangian (\ref{e1}). This puts then limits on NP contributions in $c\to u l^+ l^-$ in a model independent way. 
For  analyses of NP effects in   $D^+ \to \pi^+ \mu^+ \mu^-$  one needs matrix elements of  $\bar u_L \gamma_\mu c_L $ and 
$\bar u \sigma^{\mu \nu} (1 +\gamma_5) c$. We follow here  standard parametrisation of these matrix elements described in \cite{Fajfer:2007dy,Fajfer:2012nr}:
$<\pi (k)| \bar u \gamma^\mu (1- \gamma_5) c| D(p)> = (p+k)^\mu f_+ (q^2) + (p-k)^\mu f_ (q^2)$.   
For the $f_+(q^2)$ form factor we use Be\v cirevi\' c -Kaidalov parametrisation \cite{Becirevic:1999kt} as given in detail in \cite{Fajfer:2012nr}. 
The tensor current matrix elemet is parametrised as   $<\pi (k)| \bar u \sigma^{\mu  \nu} c| D(p)> = i {2 f_T(q^2)}/{m_D +m_\pi} [ (p+k)^\mu  q^\nu - (p+k)^\nu  q^\mu ]$. 
The HFAG report \cite{Amhis:2014hma} was used to for the relevant parameters present in $f_+(q^2)$ and $f_T(q^2)$ as given in \cite{Fajfer:2012nr}.
Based on the effective Lagrangian (\ref{e1}), the most general expression for the short distance amplitude can be written as:
 \begin{eqnarray}
 &&{\cal M}_{SD} (D^+(p) \to \pi ^+(k) \mu^+(p_+) \mu^-(p_-)) 
=\frac{G_F}{\sqrt{2}} \lambda_b  \alpha \{\biggl[ \frac{ m_c}{m_D+ m_\pi} \frac{4}{\pi} C_7   f_T(q^2)\nonumber\\
&&+ \frac{1}{\pi} C_9  f_+(q^2) \biggl] \bar u(p_-) p_\alpha \gamma^\alpha  v(p_+) 
 + \frac{1}{\pi} C_{10} f_+(q^2)  \bar u(p_-) p_\alpha \gamma^\alpha \gamma_5 v(p_+)  \}\, .
 \label{ADpi}
 \end{eqnarray}
The branching ratio for $D\to \mu^+ \mu^-$ can be written as:
 \begin{eqnarray}
&&BR(D^0 \to \mu^+ \mu^-)=   \frac{1}{\Gamma_D} = \frac{G_F^2 \alpha^2}{ 64 \pi^3} |V_{cb}^* V_{ub} |^2 f_D^2 m_D^3 
\sqrt{1 -\frac{4 m_\mu^2}{m_D^2}  }|\frac{2 m_\mu^2}{m_D^2} C_{10} |^2\, .
\label{BrD}
 \end{eqnarray}
Within SM long distance dynamics can be described be the processes  $D^+ \to \pi^+ V^0$ with $V^0 \rho^0$, $ \omega$ and $\phi$ in which  then $V^0 $ decays to $  \mu^+ \mu^-$  pair, presented  in details  in Ref. \cite{Fajfer:2007dy} for the contribution of  $D^+ \to \pi^+ \rho^0 (\omega)$ and updated for the $D^+ \to \pi^+ \phi \to \pi^+  \mu^+ \mu^-$  in Ref. \cite{Fajfer:2012nr}. 
The existing experimental upper bound in the non-resonance regions indicates that the long distance contribution is fairly suppressed. 
On Fig. 1 we present SM contributions to the differential branching fraction for $D^+ \to \pi^+ \mu^+ \mu^-$  as a function of dilepton invariant mass. We also give experimental  upper bound for the differential branching ratio as found by LHCb  \cite{Aaij:2013sua}.
\begin{figure}[htb]
\centering
\includegraphics[height=2.5in]{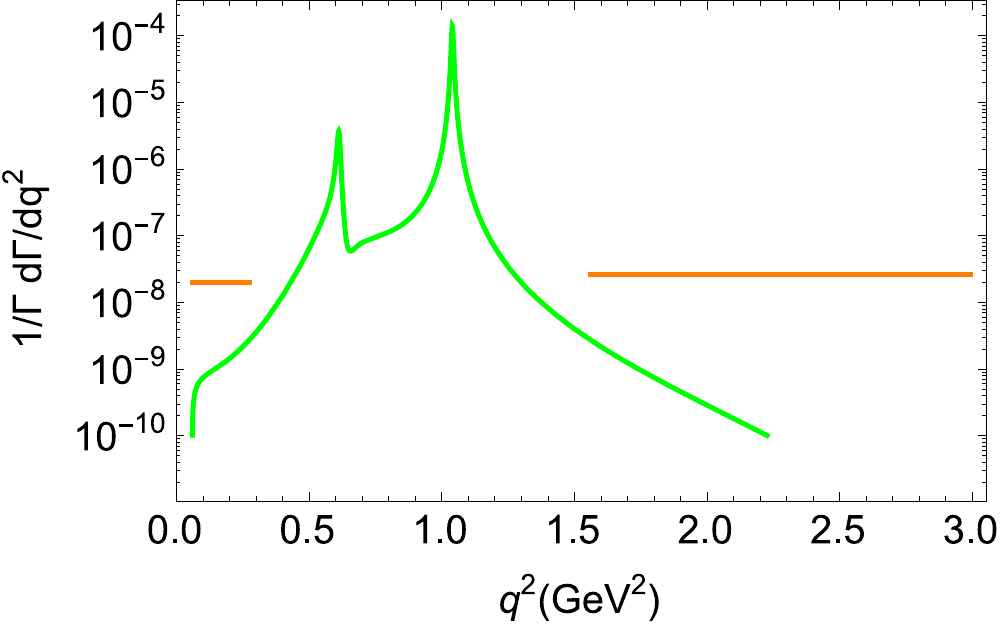}
\caption{Long distance contributions to differential branching ratio for $D^+ \to \pi^+ \mu^+ \mu^-$, as a function of dilepton invariant mass (green) and the LHCb bounds for the non-resonantt bins (orange). }
\label{fig:magnet}
\end{figure}

If one considers contributions  of  NP and if new particle is a new scalar or pseudoscalar particle mediating the decay $c\to u   \ell^+ \ell^-$, then the same particle would contribute to $D^0 - \bar D^0$ oscillations  and the physical observable from this process restricts the couplings of this operator. The same holds for the flavor changing $Z$  or new $Z^\prime$ boson. 
In the case that  NP is generated at the loop level in $c\to u   \ell^+ \ell^-$ then it contributes to $D^0  - \bar D^0$ at the loop level too, as presented in Fig. 2.
In addition to differential branching ratios at low/high dilepton invariant mass NP  detection in these decays was also discussed by suggesting new observables.  It was found that  two angular asymmetries, namely the T-odd di-plane asymmetry and the forward-backward dilepton asymmetry offer direct tests of new physics due to tiny SM backgrounds \cite{Paul:2011ar}.

\begin{figure}
\begin{center}
\includegraphics[height=2.5in]{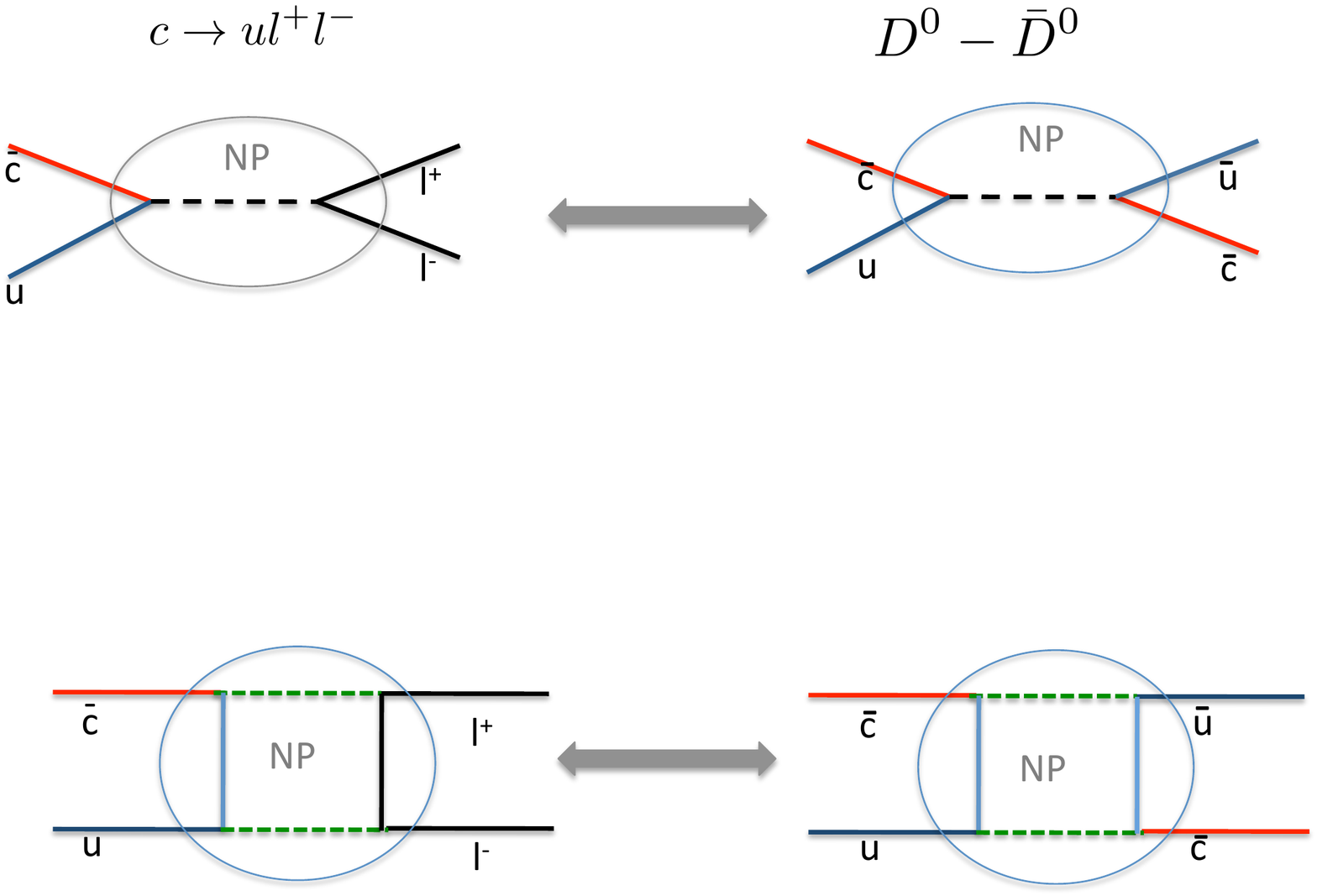}
\end{center}
\caption{\label{label}NP contributions to $c\to u l^+ l^-$ and $D^0 - \bar D^0$ at tree and loop level.}
\end{figure}

It is important to mention that outside resonance  regions of $D^+ \to \pi^+ \mu^+ \mu^-$ the long distance, as well as SM short distance contributions are more than two orders of magnitude smaller than the total branching ratio.  Experimental results for the differential decay width distribution at the low/high   dilepton invariant mass bins can be explained by the contributions of the effective Wilson coefficients. 
We allow only one Wilson coefficient at the time to have maximal value.  At the same time $D^0 \to \mu^- \mu^-$ can give bound on the $C_{10}$. Results are presented in Table \ref{tab:BR}. It turned out that  the upper bound on $D^0\to \mu^+ \mu^-$ is more restrictive on $C_{10}$ Wilson coefficient than any of the differential branching ratios for $D^+ \to \pi^+ \mu^+ \mu^-$ in dilepton  invariant mass bins. In Table \ref{tab:BR} we use  notation 
$\tilde C_i = V_{ub}  V_{cb}^* C_i$.
\begin{table}[t]
\begin{center}
\begin{tabular}{|l| |c |c|c|}  
 \hline
Max. $\tilde C_i $ & $(D\to \pi \mu \mu)_{l.i.b.}$ &  $(D\to \pi \mu \mu)_{h.i.b.}$ &  $D\to \mu \mu $ \\ 
\hline \hline
$|\tilde C_7|$ & 0.54 & 0.46 & - \\
$|\tilde C_9|$ & 1.33 & 0.91 & - \\
$|\tilde C_{10}|$ & 1.32  & 0.68& 0.63\\
\hline
$|\tilde C_9|= -|\tilde C_{10}|$ & 0.91& 0.54&  -\\
 \hline
\end{tabular}
\caption{Maximally allowed value of the Wilson coefficients, $\tilde C_i = V_{ub}  V_{cb}^* C_i$, calculated in the non-resonance regions of $D^+ \to \pi^+ \mu^+ \mu^-$ at the low lepton invariant mass $ m_{\mu\mu} \in [0.25,0.525]$ GeV, denoted by $l.i.b.$ and at the  high invariant mass region $ m_{\mu\mu} \in [1.25, 2.0]$ GeV, denoted by $h.i.b.$,  and from the upper bound on the rate of $\rm{Br} (D^0 \to \mu^+ \mu^-) < 7.6 \times 10^{-9}$.  The last row gives the  maximal value for the case $|\tilde C_9|= -|\tilde C_{10}|$.}
\label{tab:BR}
\end{center}
\end{table}
The differential rate at  high dilepton invariant  mass bin is more restrictive for the effective Wilson coefficients than the differential rate at low dilepton invariant mass bin.  Many models of 
NP were discussed in literature as supersymmetry with or without R-parity violation, new vector-like quarks, leptoquarks  \cite{Fajfer:2007dy,Fajfer:2012nr}, Little Higgs models \cite{Paul:2011ar}. All of these models modify some of Wilson coefficients  $C_i$, $i=7,9, 10$, however, they are still smaller than the bounds given  in Table \ref{tab:BR}.
The detailed analysis of the  semileptonic four body $ D\to  hh l^+ l^-$ decays was done in the work  of Ref. \cite{Cappiello:2012vg}. The dominant long distance 
contributions (bremsstrahlung and hadronic effects) are calculated and  total branching ratios and  the Dalitz plots are presented.
Assuming  vector meson dominance, it was found  $BR(D^0 \to  K^- \pi^+ l^+ l^-)  \sim 10^{-5}$, $BR(D^0 \to \pi^- \pi^+ l^+ l^-)  \sim 10^{-6}$, $BR(D^0 \to K^- K^+ l^+ l^-)  \sim 10^{-7}$ and $BR(D^0 \to K^+ \pi^- l^+ l^-)  \sim 10^{-8}$.

\section{ Direct CP violation in rare D  decays }

In 2011 the  LHCb collaboration \cite{Aaij:2011in} 
found rather large CP violation in the difference of CP violating asymmetries for  $D \to \pi^+ \pi^- (K^+ K^-)$.  Using the updated result  of the LHCb  collaboration \cite{Aaij:2014gsa}, HFAG  
\cite{Amhis:2014hma} produced the world average CP asymmetry, 
$\Delta A_{CP} = A_{CP}(D \to K^+ K^-) -A_{CP}(D \to \pi^+ \pi^-)= (-0.253 \pm 0.104)\%$.  Whether CP violating asymmetry is present  in $D \to \pi^+ \pi^- (K^+ K^-)$ or not   is still an open issue.  For  charm meson decays,  
 the CP violating asymmetry is defined as:
\begin{equation}
A_{CP}= \frac{\Gamma (D^0 \to f) - \Gamma (\bar D^0 \to f) }{\Gamma (D^0 \to f)  + \Gamma (\bar D^0 \to f) }.
\label{D-CP}
\end{equation}
The authors of Ref. \cite{Isidori:2012yx} investigated CP violating observables in $D \to V \gamma  \to \ P^+ P^- \gamma $ decays. 
If CP violation is due to NP effects  \cite{Isidori:2012yx},  then it is most likely a result of 
the chromomagnetic operator $Q_8$  contribution  ~\cite{Delaunay:2012cz}. In the case of rare $D$ decays CP violation results  from mixing of $ Q_8$ into $Q_7$ under QCD renormalization. ${\rm Im} [C_7^{NP}(m_c) ]| \simeq {\rm Im} [C_8^{NP}(m_c) ]| \simeq 0.02 \times 10^{-2}$. 
NP contribution can be  comparable in size with the real part of the SM $|C_7^{SM-eff }(m_c) |= (0.5 \pm 0.1) \times 10^{-2}$. This implies that if the phase of long distance contribution can be neglected and the relative strong phase is maximal,  the CP asymmetry can reach the ${\cal O} (1\%)$ level. 
The current world average of the  CP violating asymmetry in $\Delta  A_{CP} $, 
following the work of  \cite{Isidori:2012yx} leads to a CP asymmetry in $D\to K^+ K^- \gamma$ of the order $~ 1\%$. 

We found that in   $D \to P  \ell^+ \ell^-$  there is a  possibility  to study CP violation observables  \cite{Fajfer:2012nr}. 
 New CP violating
  effects in rare decays $D \to P \ell^+ \ell^-$ are consequence  of 
  the interference of resonant part of the long distance contribution
  and the new physics affected short distance contribution. 
    The observables, 
  the differential direct CP asymmetry and partial decay width CP
  asymmetry can be introduced  in a model independent way.  
  Among all decay modes the simplest one for the experimental searches are $D^+ \to \pi^+
  \ell^+ \ell^-$ and $D_s^+ \to K^+ \ell^+ \ell^-$.  Only when  third
generation is included  there is a possibility to obtain  non-vanishing imaginary
part: $\Im (\lambda_{b}/\lambda_d) = -\Im
(\lambda_{s}/\lambda_d)$. 
The CP violating parts of the
amplitude are suppressed by a very small factor $\lambda_b/\lambda_d \sim
10^{-3}$ with respect to the CP conserving ones and   therefore the CP violating effects should be very small. 
 In vicinity of  the $\phi$ resonant peak, the long distance amplitude for  $D^+ \to \pi^+ \mu^+ \mu^-$ decay is well 
 approximated by non-factorizable contributions of
four-quark operators in $\mc{H}^s$. The width of $\phi$ resonance is
very narrow ($\Gamma_\phi/m_\phi \approx 4\E{-3}$) and well separated
from other vector resonances in the $\qq$ spectrum of $D \to P \ell^+
\ell^-$. Relying on vector meson dominance hypothesis the
$q^2$-dependence of the decay spectrum close to the resonant peak
follows the Breit-Wigner
shape~\cite{Burdman:2001tf,Fajfer:2007dy}. 
With $\mc{A}_{LD}^\phi
= \bar{\mc{A}}_{LD}^\phi$ the differential direct CP violation becomes
\begin{eqnarray}
  \label{e16}
  a_{CP} (\sqrt{q^2}) &\equiv&
  \frac{|\mc{A}|^2-|\overline{\mc{A}}|^2}{|\mc{A}|^2+|\overline{\mc{A}}|^2}
\sim  \Im\left[ \frac{\lambda_b}{\lambda_s} C_7\right]\,.
\end{eqnarray}
 The asymmetry  can reach $a_{CP }\sim 1\%$ (see discussion in \cite{Fajfer:2012nr}).
 In addition, a CP asymmetry of a partial width in the range
$m_1< m_{\ell\ell} < m_2$ can be defined as:
\begin{equation}
\label{e17a}
  A_\mrm{CP} (m_1,m_2) = 
  \frac{\Gamma(m_1< m_{\ell\ell} < m_2)-\bar \Gamma(m_1 < m_{\ell\ell} < m_2)}{\Gamma(m_1< m_{\ell\ell} < m_2)+\bar \Gamma(m_1 < m_{\ell\ell} < m_2)}\,,
\end{equation}
where $\Gamma$ and $\bar \Gamma$ denote partial decay widths of $D^+$
and $D^-$ decays, respectively, to $\pi^\pm \mu^+ \mu^-$.  $A_\mrm{CP}$
can be related to the differential asymmetry $a_\mrm{CP}(\sqrt{q^2})$ as described in \cite{Fajfer:2012nr}.
The largest possible asymmetries are of the order few percent. 
The   new physics detection in these decay modes was also discussed. It was found that  two angular asymmetries, namely the T-odd diplane asymmetry and the forward-backward dilepton asymmetry offer direct tests of New Physics due to tiny Standard model backgrounds.  If supersymmetric and $Z^\prime$-enhanced scenarios are assumed,  and if the size of Wilson coefficients   $C_9$  and $C_{10}$  is compatible with the observed CP asymmetry in nonleptonic charm decays  and flavor constraints, it was found in \cite{Cappiello:2012vg} that new physics effects in $ D^0 \to h_1 h_2 l^+l^-$ might  reach the percent  level.

\begin{table}[t]
\begin{center}
\begin{tabular}{| l |c |c|}  
 \hline
Decay mode &   size &Reference \\ 
\hline
 $D\to \rho(\omega)\gamma$  &   $\le 3\%$  &   \cite{Lyon:2012fk}\\
  $D\to K^+ K^- \gamma$ & $\le 1\%$ &    \cite{Isidori:2012yx}  \\  
 $D\to X_u l^+ l^-$  & $\le3\%$ &  \cite{Paul:2012ab}   \\
 $D^+ \to \pi^+ \mu^+ \mu^-$  &  $\le 2\%$ & \cite{Fajfer:2012nr} \\
 $D^+ \to h h \mu^+ \mu^-$  &  $\le 1\%$&    \cite{Cappiello:2012vg} \\
 \hline
\end{tabular}
\caption{CP violating asymmetries for charm rare decays, the size  and the original reference.}
\label{tab:CP}
\end{center}
\end{table}

   If supersymmetric and $Z^\prime$-enhanced scenarios are assumed  and if the size of Wilson coefficients   $C_9$  and $C_{10}$  is compatible with the observed CP asymmetry in nonleptonic charm decays  and flavor constraints, it was found in \cite{Cappiello:2012vg} that new physics effects in $ D^0 \to h_1 h_2 l^+l^-$ might  reach the $\sim 1\%$ level.

\section{Summary}
Within SM rare charm decays are  fully dominated by long distance dynamics. 
Recent results of LHCb experiment on $D\to \mu^+ \mu^-$ and $D^+ \to \pi^+ \mu^+ \mu^-$ enable to determine bounds on effective Wilson coefficients: $C_i$, $i=7,\, 9,\,10$. 

It was found that some signals of new physics might arise in  $D\to K^+K^-\gamma$, as well as in decays with the leptonic pair in the final state 
$D\to X_u l^+ l^-$, $D^+ \to \pi^+ \mu^+ \mu^-$, $D^+ \to h h \mu^+ \mu^-$.  
In discussion of NP in rare charm decays it is necessary to check whether $D^0 - \bar D^0$ oscillations give additional constraints on the couplings of new physics. At the same time, K and B physics might for doublets of up-like quarks interacting with NP particle, give very restrictive bounds. 

The possible presence of CP violation induced by new physics  in charm nonleptonic decays 
has stimulated  a number of studies. 
The three body $D \to P \ell^+ \ell^-$  decay decays is particularly interesting, since one can focus on the CP asymmetry around the $\phi$ resonant peak in
spectrum of dilepton invariant mass.  The appropriate observables, 
  the differential direct CP asymmetry and the partial decay width CP
  asymmetry can be introduced  in a model independent way. 
  
Although long distance dynamics overshadows short distance contributions in rare charm decays, more precise measurements and improved knowledge of hadronic quantities, might uncover presence of New Physics.





\end{document}